\documentclass[a4paper,conference,10pt]{IEEEtran}

\usepackage{amsmath,epsfig, amsthm, amssymb, enumerate,algorithmic,algorithm}

\usepackage{graphicx}
\usepackage{subfigure}
\usepackage{amsmath, amssymb, epsfig}
\usepackage{bm}
\usepackage{mathrsfs}
\usepackage{color}
\usepackage{epstopdf}

\def \eps {\varepsilon}
\DeclareMathOperator*{\argmin}{arg min} 
 
\DeclareMathOperator*{\supp}{supp}

\newcommand{\R}{\mathbb{R}}

\newcommand{\C}{\mathbb{C}}

\newcommand{\oper}[1]{\mathcal{#1}}
\newcommand{\inner}[1]{\left<#1\right>}

\newcommand{\lp}[1]{\langle}
\newcommand{\rp}[1]{\rangle}

\newcommand{\diag}{\operatorname{diag}}
\newcommand{\norm}[1]{\left\|#1\right\|}

\renewcommand{\supp}{\mbox{supp}\,}

\newcommand{\beq}{\begin{equation}}
\newcommand{\eeq}{\end{equation}}
\renewcommand{\tilde}{\widetilde}
\renewcommand{\hat}{\widehat}
\newcommand{\bit}{\begin{itemize}}
\newcommand{\eit}{\end{itemize}}
\newcommand{\ben}{\begin{enumerate}}
\newcommand{\een}{\end{enumerate}}
\newcommand{\Ran}{\mbox{Ran}\,}

\newtheorem{theorem}{Theorem}[]
\newtheorem{lemma}[theorem]{Lemma}
\newtheorem{proposition}[theorem]{Proposition}

\begin{document}

\title{Super-resolution via \\ superset selection and pruning}

\author{\IEEEauthorblockN{Laurent Demanet}
\IEEEauthorblockA{Department of Mathematics\\
Massachusetts Institute of Technology\\
Cambridge, MA 02139\\
Email: laurent@math.mit.edu}
\and
\IEEEauthorblockN{Deanna Needell}
\IEEEauthorblockA{Department of Mathematics\\
Claremont McKenna College\\
Claremont, CA 91711\\
Email: dneedell@cmc.edu}
\and
\IEEEauthorblockN{Nam Nguyen}
\IEEEauthorblockA{Department of Mathematics\\
Massachusetts Institute of Technology\\
Cambridge, MA 02139\\
Email: namnguyen@math.mit.edu}}


\maketitle

\begin{abstract}
We present a pursuit-like algorithm that we call the ``superset method" for recovery of sparse vectors from consecutive Fourier measurements in the super-resolution regime. The algorithm has a subspace identification step that hinges on the translation invariance of the Fourier transform, followed by a removal step to estimate the solution's support. The superset method is always successful in the noiseless regime (unlike $\ell_1$ minimization) and generalizes to higher dimensions (unlike the matrix pencil method). Relative robustness to noise is demonstrated numerically.
\end{abstract}

\medskip

{\bf Acknowledgments.} LD acknowledges funding from the Air Force Office of Scientific Research, the National Science Foundation, and the Alfred P. Sloan Foundation. LD is grateful to Jean-Francois Mercier and George Papanicolaou for early discussions on super-resolution.

\section{Introduction}


We consider the problem of recovering a sparse vector $x_0 \in \R^n$, or an approximation thereof, from $m \leq n$ contiguous Fourier measurements 
\begin{equation}\label{eq:P}
y = A x_0 + e, 
\end{equation}
where $A$ is the partial, short and wide Fourier matrix $A_{jk} = e ^{2\pi i j k/n}$, $0 \leq j < m$, $-n/2 \leq k < n/2$, $n$ even, and, say, $e \sim N(0,\sigma^2 I_m)$.

When recovery is successful in this scenario of contiguous measurements, we may speak of super-resolution: the spacing between neighboring nonzero components in $x_0$ can be much smaller than the Rayleigh limit $n/m$ suggested by Shannon-Nyquist theory. But in contrast to the compressed sensing scenario, where the $m$ values of $j$ are drawn at random from $\{ 0, \ldots, n-1 \}$, super-resolution can be arbitrarily ill-posed. Open questions concern not only recovery bounds, but the very algorithms needed to define good estimators.

Various techniques have been proposed in the literature to tackle super-resolution, such as MUSIC \cite{schmidt1986MUSIC}, Prony's method / finite rate of innovation \cite{grenander1958toeplitz} \cite{adamjan1971analytic} \cite{vetterli2002sampling}, the matrix pencil method \cite{hua1990matrix}, $\ell_1$ minimization \cite{fuchs2005superresolution} \cite{donoho2005nonnegativeL1} \cite{castro2012superresolution} \cite{candes2012towards}, and greedy pursuits \cite{fannjiang2012CS}. 

Prony and matrix pencil methods are based on eigenvalue computations: they work well with exact measurements, but their performance is poorly understood in the presence of noise, and they are not obviously set up in higher dimensions. As for $\ell_1$ minimization, there is good evidence that $k$-sparse \emph{nonnegative} signals can be recovered from only $2k+1$ noiseless Fourier coefficients by imposing the positivity constraint with or without $\ell_1$ minimization, see \cite{donoho1992maximum} \cite{fuchs2005superresolution} and \cite{donoho2005nonnegativeL1}.  The work of \cite{castro2012superresolution} extends this result to the continuous setting by using total variation minimization. Recently, Cand\`es and Fernandez-Granda 
showed that the solution to an $\ell_1$-minimization problem with a $\norm{A^*(y-Ax)}_1$ misfit will be close to the true signal, assuming that locations of any two consecutive nonzero coefficients are separated by at least four times the super-resolution factor $n/m$ \cite{candes2012towards}. Such optimization ideas have the advantage of being easily generalizable to higher dimensions. On the flip side, $\ell_1$ minimization super-resolution is known to fail on sparse signals with nearby components that alternate signs. 



In this paper, we discuss a simple  algorithm for solving (\ref{eq:P}) based on
\bit 
\item subspace identification as in the matrix pencil method, but without the subsequent eigenvalue computation; and
\item a removal procedure for tightening the active set, remindful of a step in certain greedy pursuits.
\eit
This algorithm can outperform the well-known matrix pencil method, as we show in the numerical section, and it is generalizable to higher dimensions. It is a one-pass procedure that does not suffer from slow convergence in situations of high coherence. We also show that the algorithm provides perfect recovery for the (not combinatorially hard in the Fourier case) noiseless $\ell_0$ problem 
\beq\label{eq:ell-zero}
\min_x | \supp x | \qquad \mbox{s.t.} \qquad Ax=y.
\eeq



\section{Noiseless subspace identification}
\label{sec::noiseless}

For completeness we start by recalling the classical uniqueness result for (\ref{eq:ell-zero}).
\begin{lemma}
Let $x_0 \in \R^n$ with support $T$ such that $m \geq 2 | T |$, and let $y = Ax_0$. Then the unique minimizer of (\ref{eq:ell-zero}) is $x_0$.
\end{lemma}

We make use of the following notations. Denote $\supp x_0$ by $T$, and write $A_T$ for the restriction of $A$ to its columns in $T$. Let $T^c$ for the complement of $T$. Let $a_k$ for the $k$-th column of $A$. The superscript $L$ is used to denote a restriction of a matrix to its first $L$ rows, as in $A^L_T$.

The ``superset method" hinges on a special property that the partial Fourier matrix $A$ does not share with arbitrary dictionaries: each column $a_k$ is \emph{translation-invariant} in the sense that any restriction of $a_k$ to $s \leq m$ consecutive elements gives rise to the same sequence, up to an overall scalar. In other words, exponentials are eigenfunctions of the translation operator. This structure is important. There is an opportunity cost in ignoring it and treating (\ref{eq:P}) as a generic compressed sensing problem.

A way to leverage translation invariance is to recognize that it gives access to the \emph{subspace} spanned by the atoms $a_k$ for $k \in T$, such that $y = \sum_{k \in T} (x_0)_k a_k$. Algorithmically, one picks a number $1 < L < m$ and juxtaposes translated copies of (restrictions of) $y$ into the Hankel matrix $Y = \mbox{Hankel}(y)$, defined as
\[
Y = \left(
           \begin{array}{ccccc}
             y_0 & y_1 &  \cdots & y_{m-L-1} \\
             y_1 & y_2 &  \cdots & y_{m-L} \\
             \vdots & \vdots & \vdots & \vdots  \\
             y_{L-1} & y_{L} & \cdots & y_{m} \\
           \end{array}
         \right).
\]
The range of $Y$ is the subspace we seek.

\begin{lemma}\label{teo:ran}
If $L \geq |T|$, then the rank of $Y$ is $|T|$, and 
\[
\Ran Y = \Ran A_T^L.
\]
\end{lemma}
%

The lemma suggests a simple recovery procedure in the noiseless case: loop over all the candidate atoms $a_k$ for $-n/2 \leq k < n/2$ and select those for which the angle
\beq\label{eq:sub-1}
\angle (a^L_k, \Ran Y) = 0.
\eeq
Once the set $T$ is identified, the solution is obtained by solving the determined system 
\beq\label{eq:sub-2}
A_T x_T = y, \qquad x_{T^c} = 0.
\eeq
This procedure (unsurprisingly) provides a solution to the noise-free $\ell_0$ sparse recovery problem (\ref{eq:ell-zero}).

\begin{theorem}\label{teo:main}
Let $x_0 \in \R^n$ with support $T$ such that $m > 2 | T |$, and let $y = Ax_0$. Consider $x$ defined by (\ref{eq:sub-1}) and (\ref{eq:sub-2}), where the Hankel matrix $Y$ is built with $|T|+1 \leq L \leq m-|T|-1$. Then $x = x_0$.
\end{theorem}

The proofs of lemma \ref{teo:ran} and theorem \ref{teo:main} hinge on the fact that $A$ has full spark.

The idea of subspace identification is at the heart of a different method, the matrix pencil, which seeks the rank-reducing numbers $z$ of the pencil
\[
\overline{Y} - z \underline{Y},
\]
where $\overline{Y}$ is $Y$ with its first row removed, and $\underline{Y}$ is $Y$ with its last row removed. These numbers $z$ are computed as the generalized eigenvalues of the couple $(\underline{Y}^* \overline{Y}, \underline{Y}^* \underline{Y})$. $z$ can also be found via solving the eigenvalues of the matrix $\underline{Y}^\dagger \overline{Y}$. When $|T| \leq L \leq m-|T|$, the collection of these generalized eigenvalues includes $e^{2 \pi i j k/n}$ for $k \in T$, as well as $m-L-|T|$ zeros. There exist variants that consider a Toeplitz matrix instead of a Hankel matrix, with slightly better numerical stability properties. When $L = |T|$, the matrix pencil method reduces to Prony's method, a numerically inferior choice that should be avoided in practice if possible. 


\section{Noisy subspace identification}

The problem becomes more difficult when the observations are contaminated by noise.  In this situation $\Ran A_T^L \ne \Ran Y$, though in low-noise situations we may still be able to recover $T$ from the indices of the smallest angles $\angle (a^L_k, \Ran Y)$.


\begin{proposition}
\label{prop::bound the angle between ak and Y}
Let $y=y_0 + e$ with $e \sim N(0,\sigma^2 I_m)$, and form the corresponding $L \times (m-L)$ matrices  $Y$ and $Y_0$ as previously. Denote the singular values of $Y^{m-L}_0$ by $s_{n,0}$. Then there exists positive $c_1, C_1$ and $c$, such that with probability at least $1- c_1 m^{-C_1}$,
\beq\label{eq:angle-eps}
\sin \angle (a_k^L, \Ran Y)  \leq  \, c \, \eps_1
\eeq
for all indices $k$ in the support set and
\beq\label{eq:eps1}
\eps_1 = \frac{|T|}{\norm{a^L_k}_2} \frac{\sigma \sqrt{L \log m}}{|x_{0_{\min}}|} \sqrt{\frac{|x_{0_{\max}}|}{s_{|T|,0}}}.
\eeq

\end{proposition}

\begin{proof}
Here we sketch the proof of this proposition.  We note that $a^L_k \in \Ran Y_0$ when $k$ is in the true support.  Thus
$$
\sin \angle (a^L_k, \Ran Y) = \frac{\norm{(I-\oper P_Y) a^L_k }_2}{\norm{a^L_k}_2} = \frac{\norm{\oper P_{Y^\perp} a^L_k }_2}{\norm{a^L_k}_2}.
$$

Denote the compact singular value decomposition of $A^L_T = U S^L V^*$. Recalling that $a^L_k \in \Ran Y_0$ and a well-known fact that $Y_0 = A^L_T D (A^{m-L}_T)^*$ where $D = \diag((x_0)_T)$, we can write $a^L_k = U \alpha = \sum_{i=1}^{|T|} \alpha_i u_i$. Thus,
\begin{equation}
\label{inq::bound sin a^L_k with Y-1st step}
\sin \angle (a^L_k, \Ran Y) \leq \sum_{i=1}^{|T|} |\alpha_i| \frac{\norm{\oper P_{Y^\perp} u_i }_2}{\norm{a^L_k}_2}.
\end{equation}
Next, since $Y=Y_0 + E = A^L_T D (A^{m-L}_T)^*+E$, we have $Y [ D (A^{m-L}_T)^*]^\dagger = A^L_T + E [ D (A^{m-L}_T)^*]^\dagger$ where $A^\dagger$ is the pseudo-inverse matrix of $A$. By multiplying both sides by $(\oper P_{Y^\perp} u_i )^*$, we get
$$
(\oper P_{Y^\perp} u_i )^* Y [ D (A^{m-L}_T)^*]^\dagger = (\oper P_{Y^\perp} u_i )^*\left( A^L_T +  E [ D (A^{m-L}_T)^*]^\dagger\right).
$$
Since the vector $\oper P_{Y^\perp} u_i$ is orthogonal to $\Ran Y$, the left hand side is zero.  Thus multiplying both sides by $v_i$, the $i$-th right singular vector of $A^L_T$, we have
$$
0 = (\oper P_{Y^\perp} u_i)^* A^L_T v_i + (\oper P_{Y^\perp} u_i )^* E [ D (A^{m-L}_T)^*]^\dagger v_i.
$$ 

\noindent We can see that $(\oper P_{Y^\perp} u_i )^* A^L_T v_i = (\oper P_{Y^\perp} u_i )^* s^L_i u_i = s^L_i \norm{\oper P_{Y^\perp} u_i }_2^2$ where $s^L_i$ is the $i$-th singular value of $A^L_T$. We therefore obtain
\begin{align*}
s^L_i \norm{\oper P_{Y^\perp} u_i }_2^2 &= -(\oper P_{Y^\perp} u_i )^* E [ D (A^{m-L}_T)^*]^\dagger v_i \\
&\leq \norm{(\oper P_{Y^\perp} u_i}_2 \norm{E} \norm{D^\dagger} \norm{[(A^{m-L}_T)^*]^\dagger}.
\end{align*}
This leads to the upper bound 
\begin{align}
\label{inq::bound l2 of ui on Yc}
\norm{\oper P_{Y^\perp} u_i }_2 &\leq \frac{1}{s^L_i} \norm{E} \norm{D^\dagger} \norm{[(A^{m-L}_T)^*]^\dagger} \notag\\
&= \frac{\norm{E}}{s^L_i} \frac{1}{|x_{0_{\min}}|} \frac{1}{s^{m-L}_{|T|}} ,
\end{align}
where $s^{m-L}_{|T|}$ is the smallest singular value of $A^{m-L}_T$.

Recalling that $a^L_k = U \alpha$, we have $\alpha_i = u^*_i a^L_k$. From the SVD of $A^L_T$, we see that $A^L_T (A^L_T)^* = U (S^L)^2 U^*$, so that
$$
U^* A^L_T (A^L_T)^* U^* = (S^L)^2.
$$
This identity implies that $\norm{u^*_i A^L_T}_2 = s^L_i$, and thus, $|\alpha_i| \leq s^L_i$. Combining this result with \eqref{inq::bound l2 of ui on Yc} and \eqref{inq::bound sin a^L_k with Y-1st step} yields
\begin{equation}
\sin \angle (a^L_k, \Ran Y) \leq |T| \frac{\norm{E}}{s^{m-L}_{|T|} |x_{0_{\min}}|} \frac{1}{\norm{a^L_k}_2}.
\end{equation}

Using the matrix Bernstein inequality of~\cite{tropp2012matBeinstein} one obtains that $\norm{E} \leq \sigma \sqrt{c L \log m}$ with high probability.  Finally, writing $Y^{m-L}_T$ as $Y^{m-L}_T = A^{m-L}_T D^{1/2} (D^{1/2})^* (A^{m-L}_T)^*$, we have 
\begin{align*}
s_{|T|,0} &= \min_z \frac{\norm{A^{m-L}_T D^{1/2} z}_2^2}{\norm{z}_2^2} = \min_{h}  \frac{\norm{A^{m-L}_T h}_2^2} {\norm{D^{-1/2} h}_2^2} \\
&\leq \min_h \frac{\norm{A^{m-L}_T h}_2^2} {\norm{h}_2^2 s_{\min} (D^{-1})} \leq (s^{m-L}_{|T|})^2 |x_{0_{\max}} |,
\end{align*}
which completes the proof.
\end{proof}

%

There are a few unknown quantities involving $\epsilon_1$, which can empirically be controlled. The support size $T$ can be estimated by a reasonably large constant, say $m/2$. The dynamic range of the signal can presumably be known if we know in prior the type of underlying signal of interest. The singular value $s_{|T|,0}$ of $Y^{m-L}_0$ can be replaced by that of $Y^{m-L}$ via the simple Weyl's inequality $| s_i - s_{i,0}| \leq \| \, \mbox{Hankel}(e) \|$, which can in turn be controlled as $O(\sigma \sqrt{L \log m})$ with high probability.



The subspace identification step now gathers all the values of $k$ such that 
\[
\sin \angle (a_k^L, \Ran Y) \leq c \, \eps_1.
\]
The resulting set $\Omega$ of indices is only expected to be a \emph{superset} of the true support $T$, with high probability. 


A second step is now needed to prune $\Omega$ in order to extract $T$. For this purpose, a loop over $k$ is set up where we test the membership of $y$ in $\Ran A_{\Omega \backslash k}$, the range of $A_\Omega$ with the $k$-th column removed. We are now considering a new set of angles where the roles of $y$ and $A$ are reversed: in a noiseless situation, $k \in T$ if and only if
\[
\angle (y, \Ran A_{\Omega \backslash k}) \ne 0.
\]
When noise is present, we first filter out the noise off $\Omega$ by projecting $y$ onto the range of $A_\Omega$, then estimate $k \in T$ only when the angle is above a certain threshold. It is easier to work directly with projections $\Pi$:
\[
\| \Pi_{\Omega} y - \Pi_{\Omega \backslash k} y \| = \sin \angle (\Pi_\Omega y, \Ran A_{\Omega \backslash k}) \;
\| \Pi_{\Omega} y \|.
\]
The effect of noise on the left-hand side is as follows.

\begin{proposition}
Let $y=y_0 + e$ with $e \sim N(0,\sigma^2 I_m)$. Let $\Pi_\Omega y$ be the projection of $y$ onto $\Ran A_\Omega$, and let $\Delta \Pi = \Pi_\Omega - \Pi_{\Omega \backslash k}$. Then there exists $c > 0$ such that, with high probability,
\[
| \, \| \Delta \Pi y \| - \| \Delta \Pi y_0 \| \, |  \leq c \, \eps_2,
\]
with $\eps_2 = \sigma$.
\end{proposition}

Algorithm \ref{alg:SSP} for the superset method implements the removal step in an iterative fashion, one atom at a time.





\begin{algorithm}[ht]
\caption{Superset selection and pruning} \label{alg:SSP}
\begin{algorithmic}
\STATE \textbf{input:} Partial Fourier matrix $A \in C^{m \times n}$, $y=Ax_0+e$, parameter $L$, thresholds $\eps_1$ and $\eps_2$.
%
%
\STATE \textbf{initialization:} $Y = \mbox{Hankel}(y) \in \C^{L \times (m-L)}$
\STATE \textbf{support identification}
\STATE
\begin{tabular}{ll}
\textbf{decompose:} & $\tilde{Q} \tilde{R} = Y \tilde{E}$, $\tilde{Q} \in \C^{L \times r}$ \\
\textbf{project:} & $a_k \leftarrow A_{\{ k \}}$ ( for all $k$)\\
							& $\gamma_k \leftarrow \norm{a_k - \tilde{Q} \tilde{Q}^* a_k}/\norm{a_k}$ \\
							& $\Omega = \{ k: \gamma_k \leq \eps_1 \}$
\end{tabular}

\WHILE{true}
\STATE
\begin{tabular}{ll}
\textbf{decompose:} & $QR = A_{\Omega}E$, $Q\in\C^{m\times|\Omega|}$ \\
\textbf{remove:} & $\forall k\in\Omega$: $Q_{(k)}R_{(k)} = A_{\Omega\backslash k}E_{(k)}$ \\
&$\delta_k \leftarrow \|(Q_{(k)} Q_{(k)}^* - QQ^*)y\|_2$ \\
 & $k_0 \leftarrow \argmin_{k} \delta_k$\\
 & if $\delta_{k_0} < \eps_2$, $\Omega \leftarrow \Omega\backslash k_0$\\
 & else break \\
\end{tabular}
\ENDWHILE
\STATE \textbf{output:} $\hat{x} = \argmin_x \norm{y - A_{\Omega} x}$
\end{algorithmic}
\end{algorithm}

\section{Experimental Results}

\begin{figure}[!t]
\centering
\includegraphics[width=1.5in]{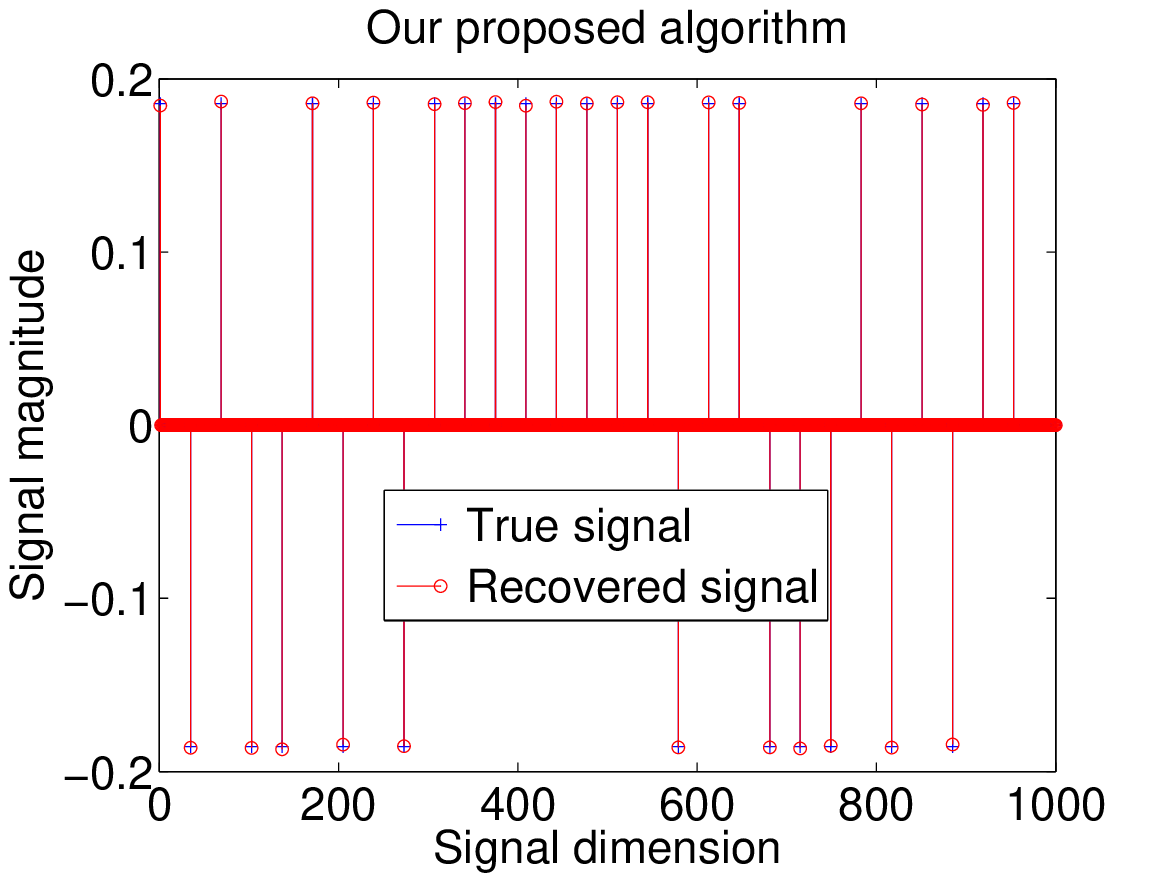} 
\hfill
\includegraphics[width=1.5in]{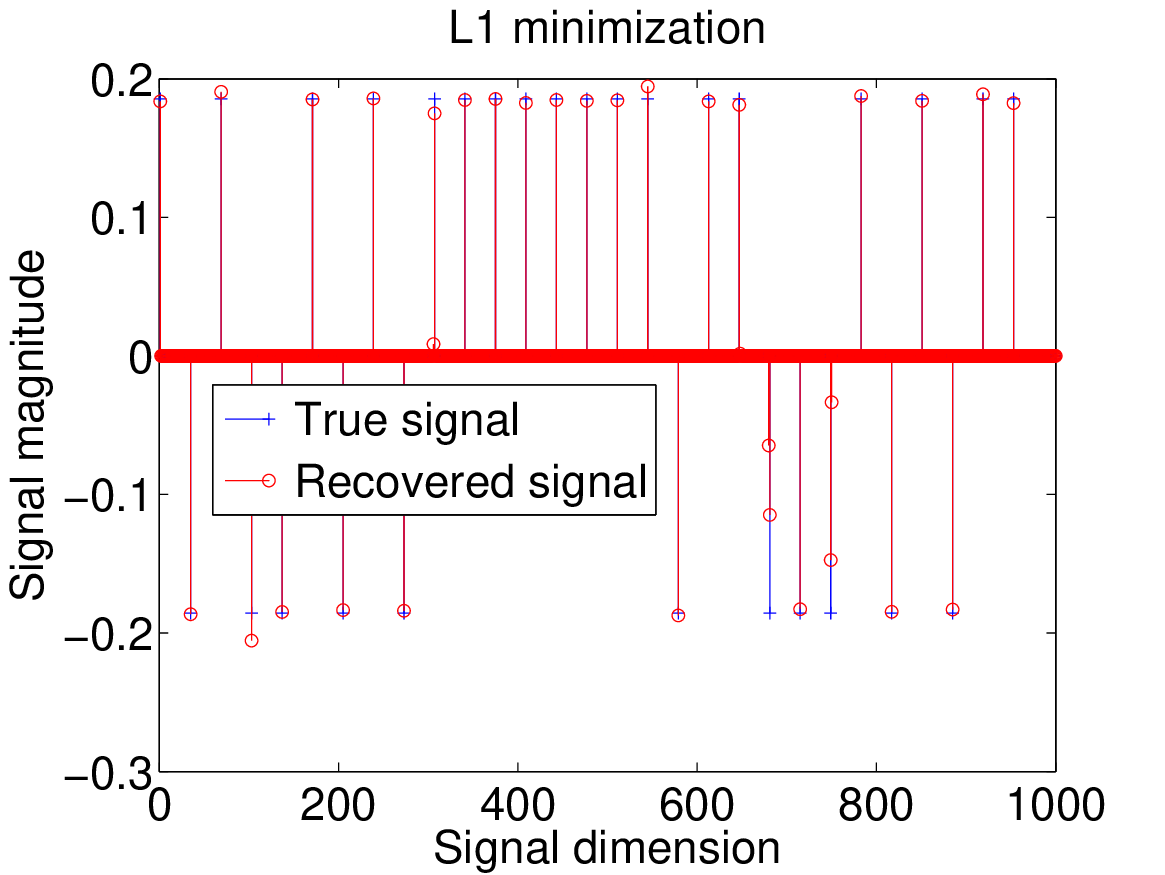} \\
\includegraphics[width=1.5in]{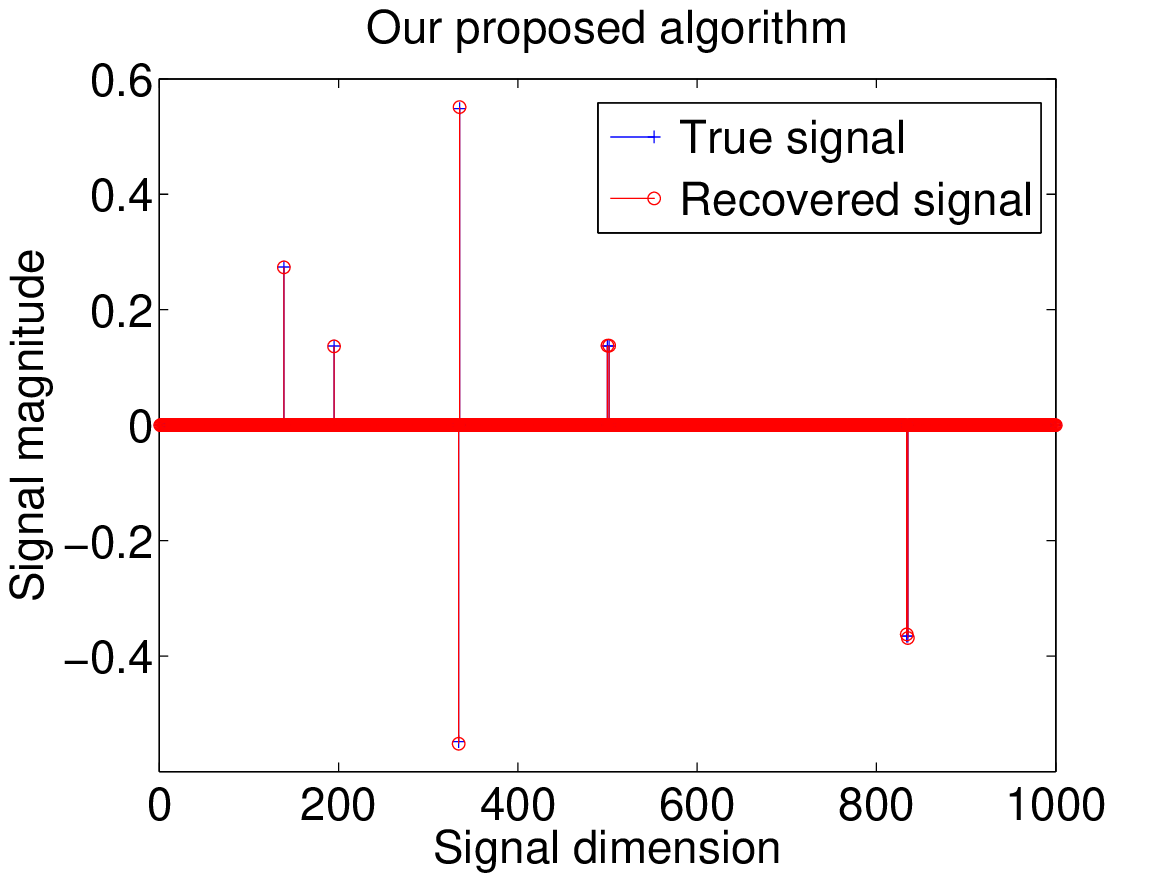}
\hfill
\includegraphics[width=1.5in]{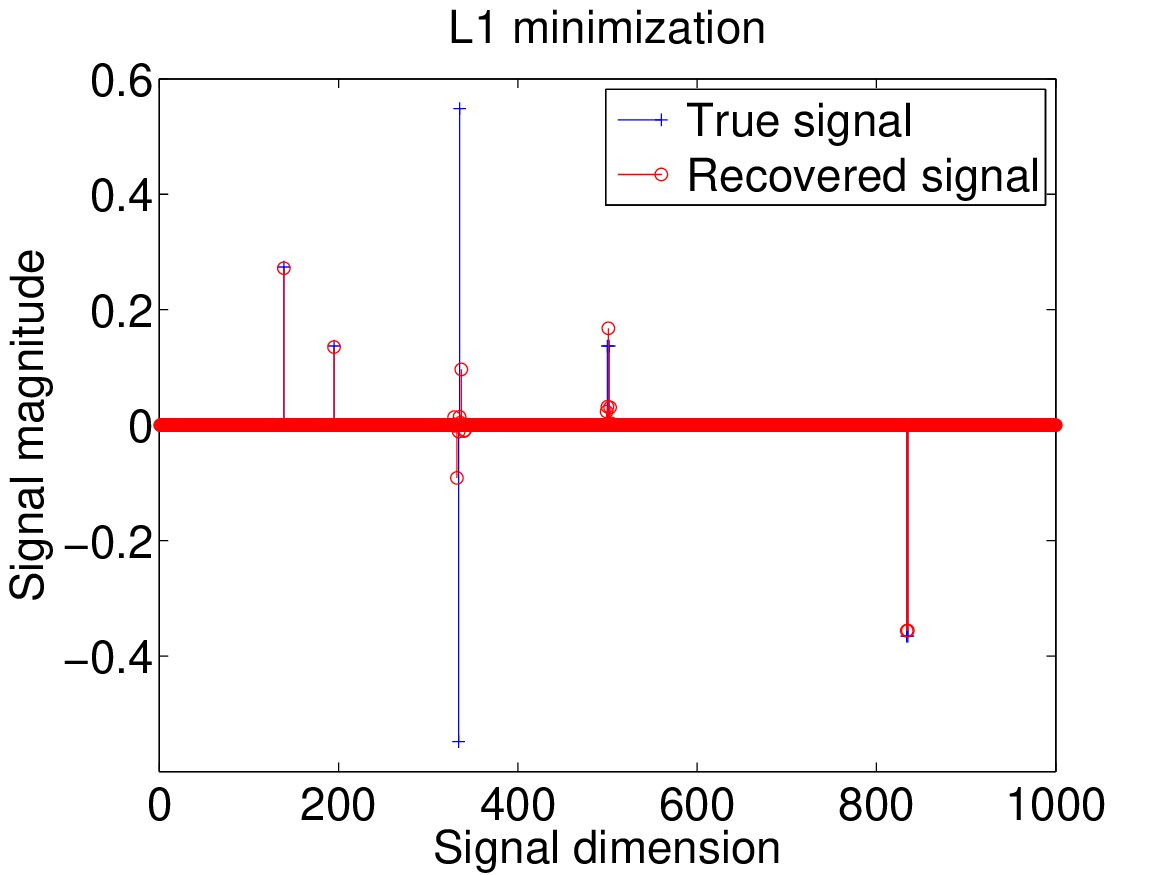} 
\caption{\small Original (blue) and recovered (red) signals. Left column: the superset method. Right column: $\ell_1$-minimization. Top row: a signal with well-separated spikes. Bottom row: spike spacing below the Rayleigh length.}
\label{fig::recovery}
\end{figure}

In the first simulation, we fix $n=1000$ and $m=120$ and construct an $n$-dimensional signal $x_0$ whose nonzero components are well separated by at least $4n/m$, a distance equivalent to four times the super-resolution factor $n/m$. The spike magnitudes are independently set to $\pm 1/\sqrt{29}$ with probability $1/2$. The noise vector $e$ is drawn from $N(0,\sigma^2 I_m)$ with $\sigma = 10^{-3}$.  We fix the thresholds $\eps_1$ via (\ref{eq:eps1}) with $c=1$ and $\eps_2 =  10 \sigma$. Throughout our simulations, we set $L = \lfloor m/3 \rfloor$.
As can be seen from Fig. \ref{fig::recovery}, top row, the recovered signal from the superset method is reasonable, with $\norm{\hat{x}-x_0}_2 = 0.075$, while the reconstruction via $\ell_1$-minimization tends to exhibit incorrect clusters around the true spikes.  


Our next simulation considers a more challenging signal model with a strongly coherent matrix $A$. For example, with $n=1000$ and $m=120$, the coherence of the matrix $A$ with normalized columns $a_i$ is $\mu = \max_{i \neq j} |\inner{a_i,a_j}| = 0.9765$.
The signal in this simulation is shown in Fig. \ref{fig::recovery}, bottom row. It consists of five spike clusters:  each of the first two clusters consists of a single spike, and each of the last four clusters contains two neighboring spikes. The signs of these neighboring spikes either agree or differ.
We set $m, \sigma$ and $\eps_2$ as in the previous simulation, and we let the constant $c$ in the equation (\ref{eq:eps1}) of  $\eps_1$ equal to 5. Recovery via the superset method is accurate, while $\ell_1$ minimization fails at least with clusters of opposite-sign spikes.

In the next simulation, we consider a signal of size $n=1000$ which contains two nearby spikes at locations $[100, 101]$ and has magnitudes $1/\sqrt{2}$ and $-1/\sqrt{2}$. We empirically investigate the algorithm's ability to recover the signal from varying measurements $m=\{10,20,...,220\}$ and noise levels $log_{10} \sigma = \{-3.5,-3.4,...,-2\}$. 
For each pair $(m,\sigma)$, we report the frequency of success over $100$ random realizations of $e$. The greyscale goes from white (100 successes) to black (100 failures).  A trial is declared successful if the recovered $\hat{x}$ satisfies $\norm{\hat{x}-x_0}_2/\norm{x_0}_2 < 10^{-3}$. 
The horizontal axis indicates the noise level $\sigma$ in log scale, and the vertical axis indicates $\log_{10} (1-\mu)$ where $\mu$ is the coherence as earlier.

\begin{figure}[htb]
\centering
\includegraphics[width=1.5in]{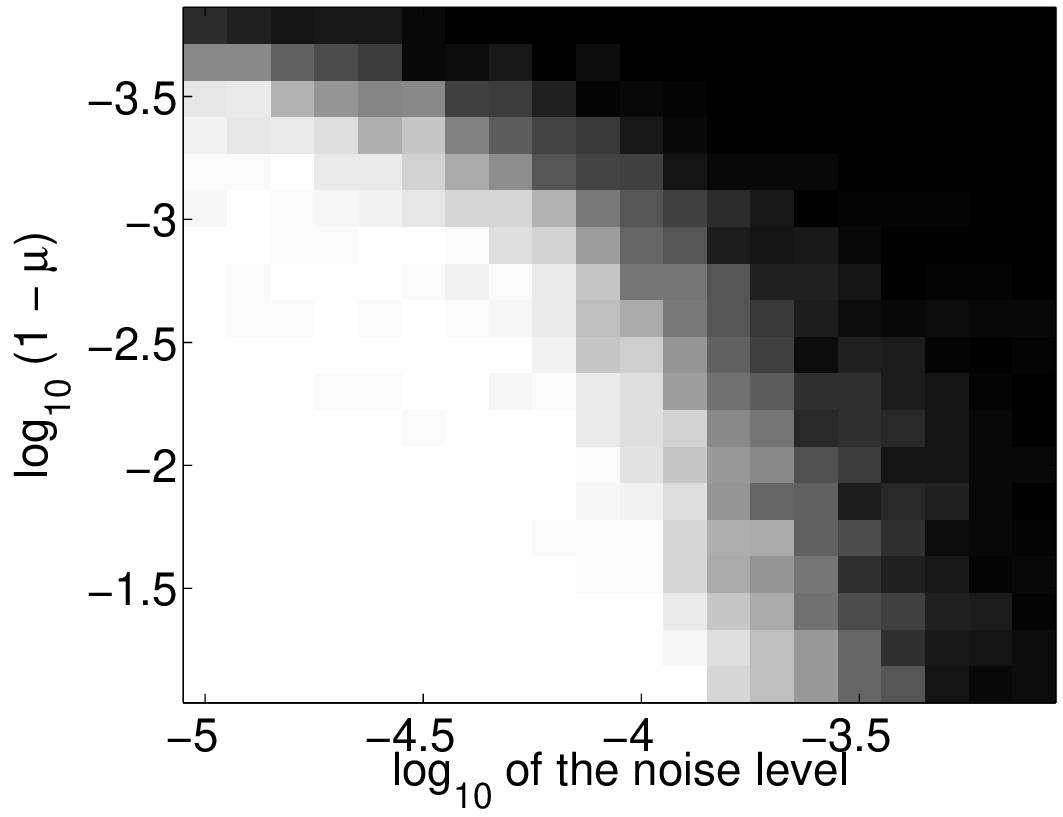} 
\hfill
\includegraphics[width=1.5in]{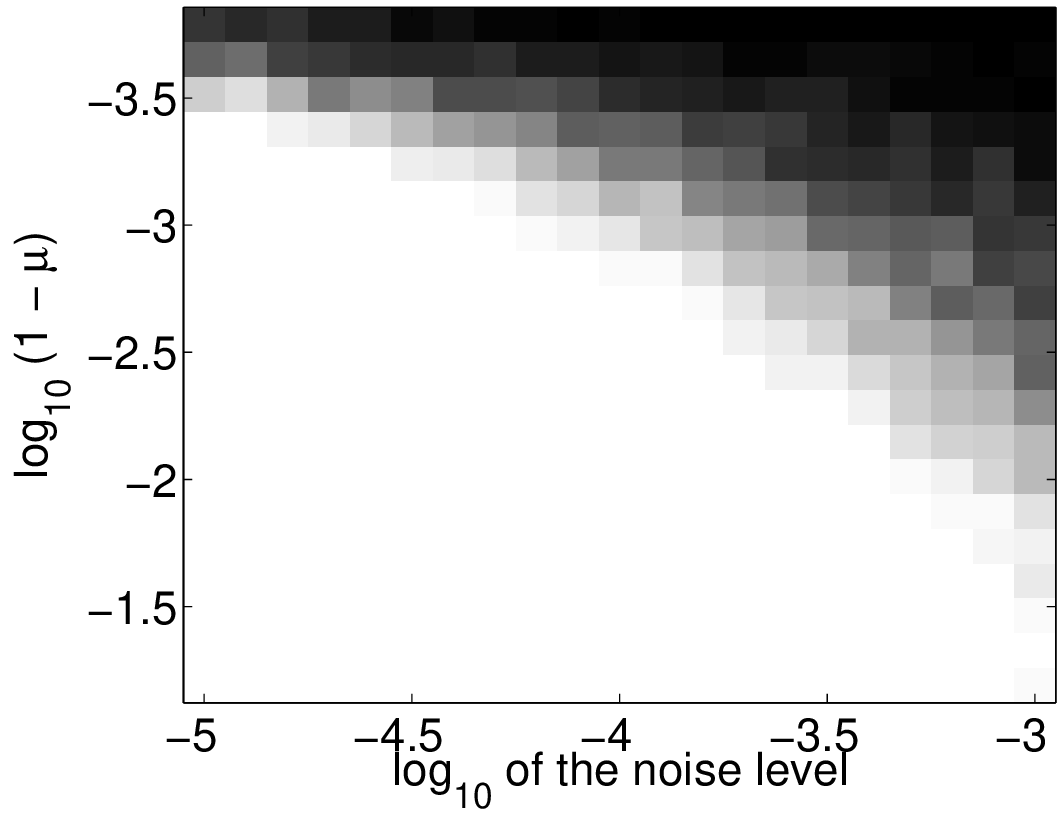} \\
\includegraphics[width=1.5in]{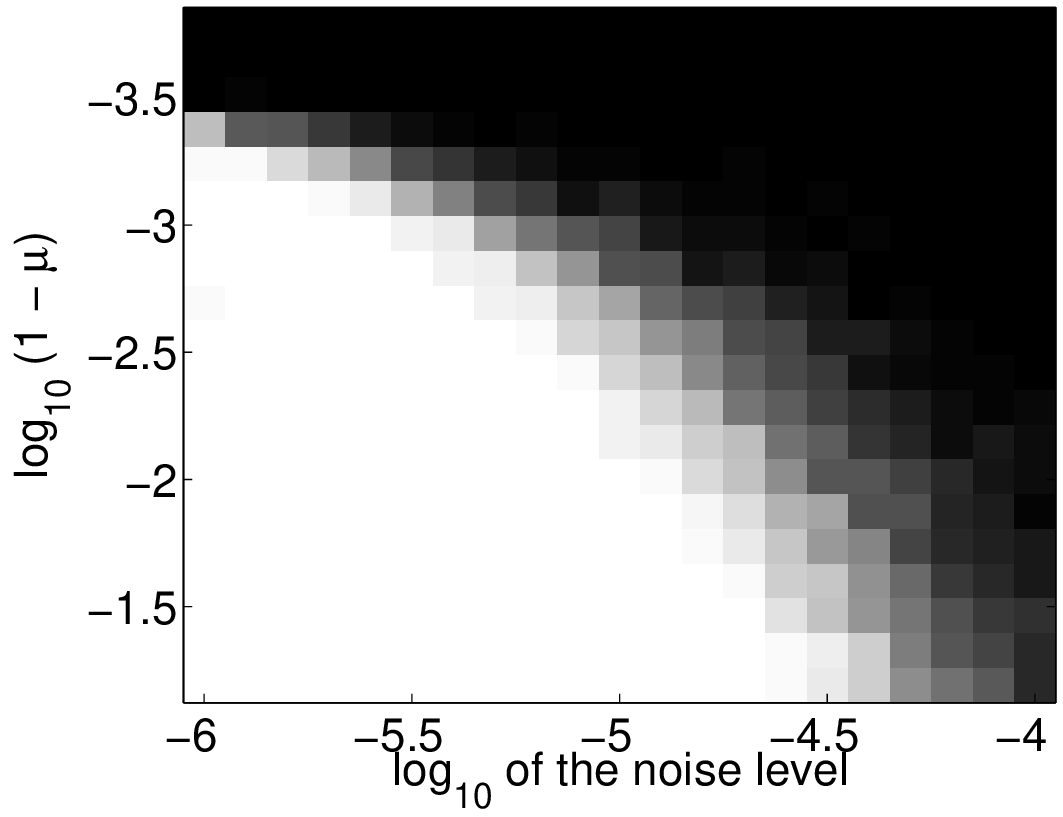} 
\hfill
\includegraphics[width=1.5in]{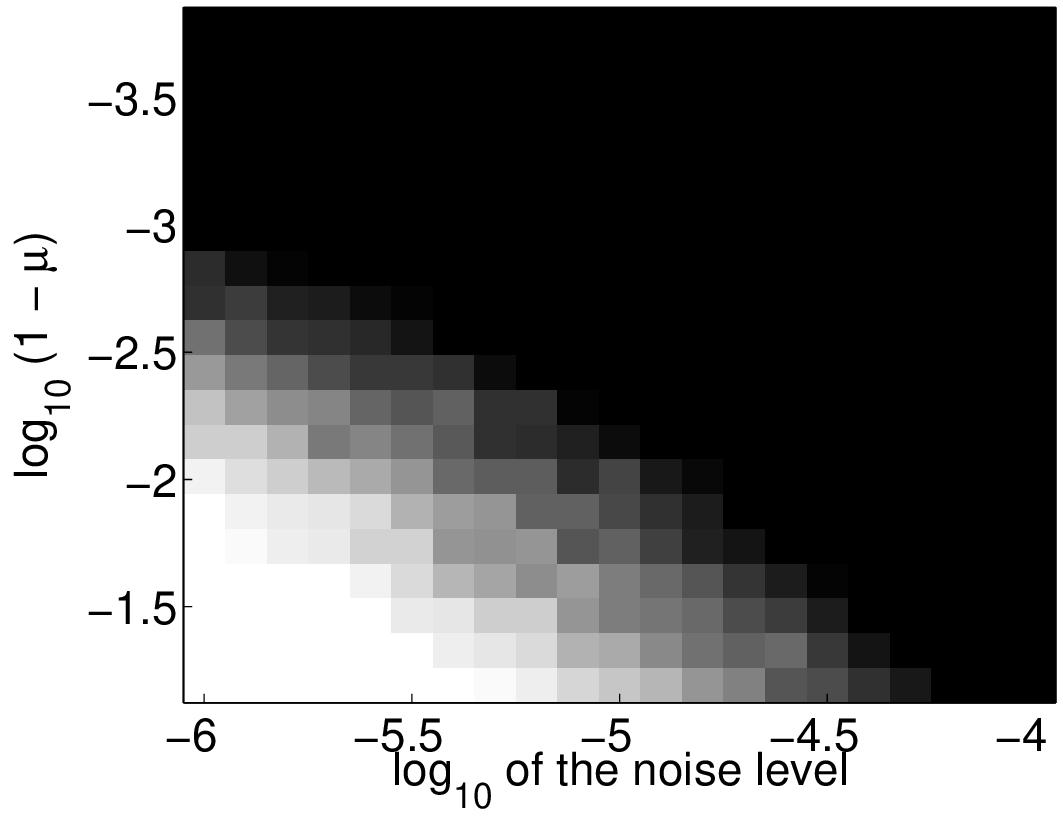} \\
\includegraphics[width=1.5in]{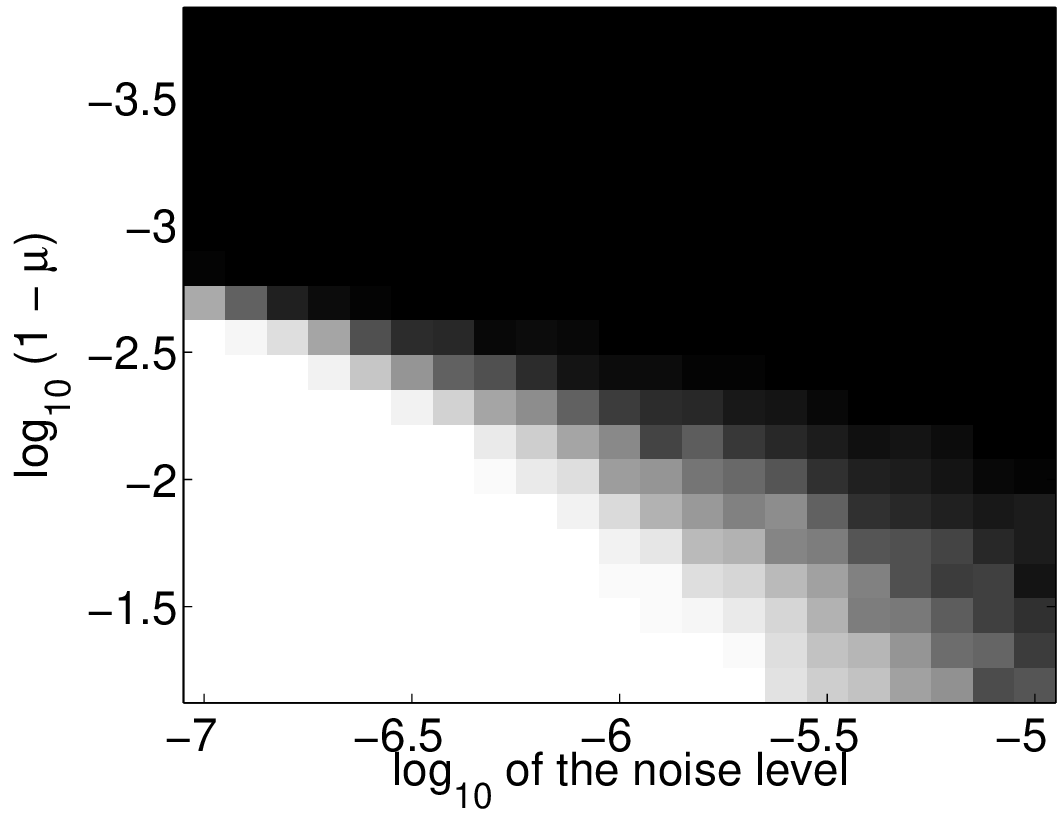} 
\hfill
\includegraphics[width=1.5in]{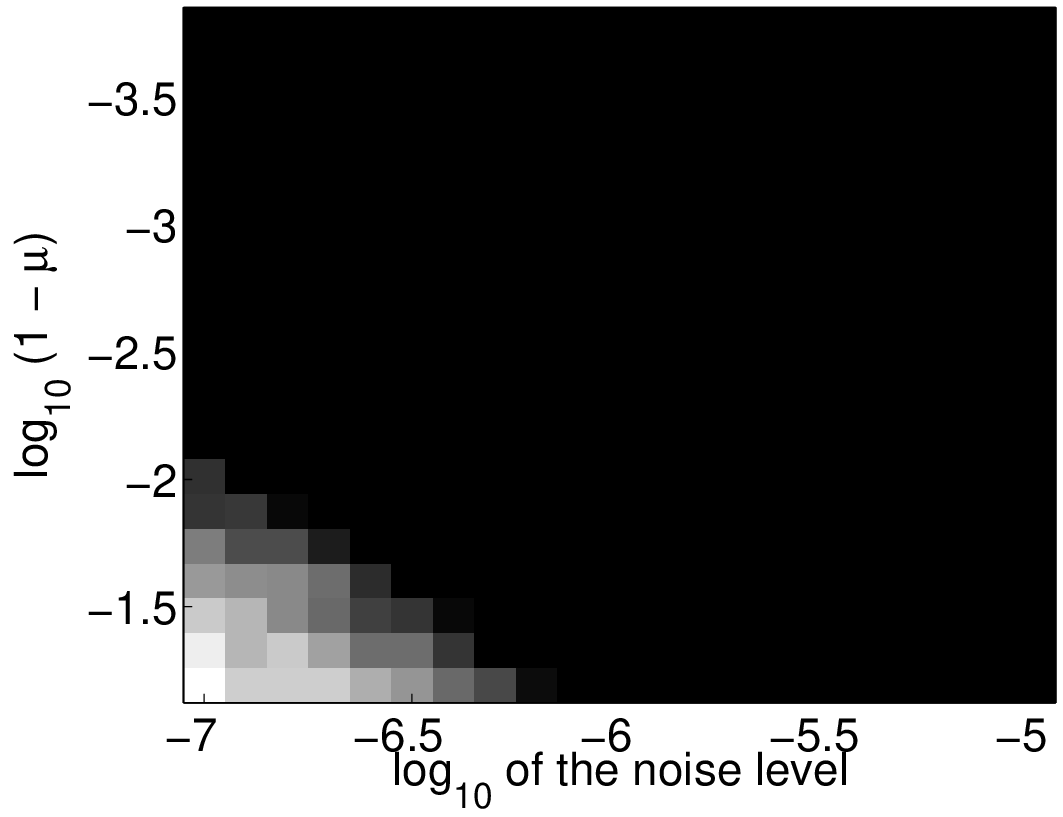}
\caption{Probability of recovery, from 1 (white) to 0 (black) for the superset method (left column) and the matrix pencil method (right column). Top row: 2-sparse signal. Middle row: 3-sparse signal. Bottom row: 4-sparse signal. The plots show recovery as a function of the noise level (x-axis, $\log_{10} \sigma$) and the coherence (y-axis, $\log_{10}(1-\mu)$).}\label{fig:phase transition}
\end{figure}

We note that the coherence is inversely proportional to the amount of measurements $m$ and proportional to the super-resolution factor $n/m$: increasing $m$ (decreasing the super-resolution factor) will reduce the coherence $\mu$. On the vertical axis, smaller values imply higher coherence, or equivalently smaller amount of measurements. As shown in Fig. \ref{fig:phase transition}, for reasonably small noise, the algorithm is able to recover the signal exactly even the coherence is nearly $1$.


For reference, we also compare the superset method with the matrix pencil method as set up in \cite{hua1991svd}. The noise is filtered out by preparing low-rank approximations of $\underline{Y}$ and $\overline{Y}$ where only the singular values above $c \sigma \sqrt{L \log L}$ are kept, for some heuristically optimized constant $c$. Two more signals are considered: (1) a 3-sparse signal consisting of three neighboring spikes, each of magnitude $1/\sqrt{3}$ with alternating signs, and (2) a 4-sparse signal with neighboring spikes of alternating signs and equal magnitude $1/2$. Fig. \ref{fig:phase transition} is a good illustration of the contrasting numerical behaviors of the two methods: the matrix pencil is often the better method in the special case of a signal with 2 spikes, but loses ground to the superset method in various cases of progressively less sparse signals. Understanding the performance of the matrix pencil would require formulating a lower bound on the (typically extremely small) $S$-th eigenvalues of $Y_0$ where $S$ is the sparsity of $y_0$.

\section{Conclusion}

Empirical evidence is presented for the potential of the superset method as a viable computational method for super-resolution. Further theoretical justifications will be presented elsewhere.

\bibliography{bib}
\bibliographystyle{plain}

\end{document}